\documentclass[journal,10pt]{IEEEtran}

\usepackage{bm}
\usepackage{amsmath,amssymb}
\usepackage{graphicx,subfigure,epstopdf}

\usepackage{color}

\newcommand{\RR}{\mathbb{R}}
\newcommand{\ZZ}{\mathbb{Z}}

\newtheorem{Assumption}{Assumption}
\newtheorem{Remark}{Remark}
\newtheorem{Lemma}{Lemma}
\newtheorem{Proposition}{Proposition}
\newtheorem{Theorem}{Theorem}

\newtheorem{Problem}{Problem}

\newenvironment{Proof}{\noindent{\em Proof:\/}}{\hfill $\Box$\par}

\begin{document}

\title{Discrete-Time Distributed Observers over Jointly Connected Switching Networks and an Application}

\author{Tao~Liu
        and~Jie~Huang
\thanks{This work has been supported by the Research Grants Council of the
Hong Kong Special Administration Region under grant No. 14200617.
\emph{(Corresponding author: Jie Huang.)}
%, and in part by the National Natural Science Foundation of China under grant No. 61174049.
}% <-this % stops a space
\thanks{Tao Liu and Jie Huang  are with
 the Department of Mechanical and Automation
Engineering, The Chinese University of Hong Kong, Shatin, N.T., Hong
Kong. E-mail: tliu2@mae.cuhk.edu.hk, jhuang@mae.cuhk.edu.hk}}

\maketitle

\begin{abstract}
In this paper, we first  establish an exponential stability result
for a class of linear switched systems and then apply this result to show the existence of the
distributed observer for a discrete-time leader system
over jointly connected switching networks.
A special case of this result
leads to the solution of a leader-following consensus problem of
multiple discrete-time double-integrator systems over jointly connected switching networks.
Then, we further develop the adaptive distributed observer for the discrete-time  leader system
over jointly connected switching networks, which
has the advantage over the distributed observer in that it does not require
that every follower know the system matrix of the leader system.
As an application of the discrete-time distributed observer,
we will solve the cooperative output regulation problem of a
discrete-time linear multi-agent system over jointly connected switching networks.
A leader-following formation problem of mobile robots
will be used to illustrate our design.
This problem cannot be handled by any existing approach.
\end{abstract}

%% Note that keywords are not normally used for peerreview papers.
%\begin{IEEEkeywords}
%IEEE, IEEEtran, journal, \LaTeX, paper, template.
%\end{IEEEkeywords}

%%%%%%%%%%%%%%%%%%%%%%%%%%%%%%%%%%%%%%%%%%%%%%%%%%%%%%%%%%%%%%%%%%%%%%%%%%%%%%%%%%%%%%%%%%%%%%%%%%%%%%%%%%%%%%%%%%%%%%%%%%%%%%%%%%%%%%%%%%%%%%%%%%%%%%%%%%%%%%%%%%%%%%%%%%%%%%%%%%%%%%%%%%%%%%%%%%

\section{Introduction}\label{Section I}
The distributed observer for a leader system is a distributed dynamic compensator that can estimate
the state of the leader distributively in a sense to be described in Section \ref{Section Three}.
It is an effective tool in dealing with various cooperative control problems for multi-agent systems.
For a linear continuous-time leader system of the following form:
\begin{equation}\label{cexo}
  \dot{v}(t)=Sv(t), \qquad    t \ge 0
\end{equation}
where $v \in \RR^{q}$ and $S \in \RR^{q \times q}$ is  a constant system matrix,
a distributed observer was first developed in \cite{SuHuang12}
over connected static networks, and then in \cite{SuHuangCyber12}
over jointly connected switching networks.
On the other hand, for a linear discrete-time leader system of the following form:
\begin{equation}\label{eq-leader-system-discrete}
  v(t+1)=Sv(t), \qquad  t =0,1,2,\ldots,
\end{equation}
a distributed observer  was also developed in \cite{YanHuang16} over connected static networks, and then in  \cite{YanHuang17}
over jointly connected switching networks.

It was shown in \cite{SuHuangCyber12} that a distributed observer for (\ref{cexo})
over jointly connected switching networks always exists if
all the eigenvalues of the matrix $S$ have zero or negative real parts.
However, the existence conditions of a distributed observer for (\ref{eq-leader-system-discrete})
over jointly connected switching networks are much more stringent than the continuous-time case.
Specifically, the result in \cite{YanHuang17} further requires that
some subgraph of the jointly connected switching digraph be undirected
and the matrix $S$ be marginally stable in the sense that
all the eigenvalues of  $S$ on the unit circle must be semi-simple.
Moreover, while the solution of each subsystem of the distributed observer for (\ref{cexo}) converges to the state of
(\ref{cexo}) exponentially,  the solution of  each subsystem of the discrete-time distributed observer for (\ref{eq-leader-system-discrete}) is only shown to converge to the state of  (\ref{eq-leader-system-discrete}) asymptotically.

Due to this significant gap between the result of the distributed observer for
the continuous-time system (\ref{cexo}) and
the result of the distributed observer for the discrete-time system (\ref{eq-leader-system-discrete}),
the applications of discrete-time multi-agent control systems also lag
far behind the applications of continuous-time multi-agent control systems.
For example, while the leader-following consensus problem of multiple continuous-time double-integrator systems over jointly connected switching networks was solved as early as 2012 in \cite{SuHuangCyber12},
the leader-following consensus problem of multiple discrete-time double-integrator  systems over jointly connected switching networks is yet to be studied.

%%%%%%%%%%%%%%%%%%%%%%%%%%%%%%%%%%%%%%%%%%%%%%%%%%%%%%%%%%%%%%%%%%%%%%%%%%%%%%%%%%%%%%%%%%%%%%%%%%%%%%%%%%%%%%%%%%%%%%%%%%%%%%%%%%%%%%%%%%%%%%%%%%%%%%%%%%%%%%%%%%%%%%%%%%%%%%%%%%%%%%%%%%%%%%%%%%

In this paper, we will further study the distributed observer for the discrete-time system (\ref{eq-leader-system-discrete}) over jointly connected switching networks.
We first  establish an exponential stability result
for a class of linear switched systems and then apply this result to show the existence of the
distributed observer for the discrete-time system (\ref{eq-leader-system-discrete}) over jointly connected switching networks. A special case of this result
leads to the solution of the leader-following consensus problem of multiple discrete-time double-integrator systems over jointly connected switching networks.
Then,  we further develop the adaptive distributed observer for the discrete-time system (\ref{eq-leader-system-discrete}) over jointly connected switching networks.
It is noted that the adaptive distributed observer for the continuous-time system (\ref{cexo})
was proposed in \cite{CaiHuang16} over jointly connected switching networks.
An advantage of the adaptive distributed observer over the distributed observer is that,
 it does not require that every follower know the system matrix $S$ of the leader.
The discrete-time version of the adaptive distributed observer over connected static networks was proposed in
\cite{Huang17} and further improved in \cite{LTHuang19}. However, the discrete-time version of the adaptive distributed observer over jointly connected switching networks is still missing.
As an application of the discrete-time distributed observer,
we will solve the cooperative output regulation problem of a
discrete-time linear multi-agent system over jointly connected switching networks.
A leader-following formation problem of mobile robots
will be used to illustrate our design. This problem cannot be handled by any
existing approach.

%%%%%%%%%%%%%%%%%%%%%%%%%%%%%%%%%%%%%%%%%%%%%%%%%%%%%%%%%%%%%%%%%%%%%%%%%%%%%%%%%%%%%%%%%%%%%%%%%%%%%%%%%%%%%%%%%%%%%%%%%%%%%%%%%%%%%%%%%%%%%%%%%%%%%%%%%%%%%%%%

The rest of the paper is organized as follows.
We present some technical lemmas in Section \ref{Section Two}
and then establish the  distributed observer and the adaptive distributed observer in Section \ref{Section Three}.
In Section \ref{Section Four}, we show the solvability of the  cooperative output regulation problem of
a discrete-time linear multi-agent system over jointly connected switching networks via the distributed observer approach.
An example is given in Section \ref{Section Five}
and the paper is concluded in Section \ref{Section Six} with some remarks.

%%%%%%%%%%%%%%%%%%%%%%%%%%%%%%%%%%%%%%%%%%%%%%%%%%%%%%%%%%%%%%%%%%%%%%%%%%%%%%%%%%%%%%%%%%%%%%%%%%%%%%%%%%%%%%%%%%%%%%%%%%%%%%%%%%%%%%%%%%%%%%%%%%%%%%%%%%%%%%%%%%%%%%%%%%%%%%%%%%%%%%%%%%%%%%%%%%

%%%%%%%%%%%%%%%%%%%%%%%%%%%%%%%%%%%%%%%%%%%%%%%%%%%%%%%%%%%%%%%%%%%%%%%%%%%%%%%%%%%%%%%%%%%%%%%%%%%%%%%%%%%%%%%%%%%%%%%%%%%%%%%%%%%%%%%%%%%%%%%%%%%%%%%%%%%%%%%%%%%%%%%%%%%%%%%%%%%%%%%%%%%%%%%%%%

\medskip

\noindent {\em Notation.} $\mathbb{Z}^+$ denotes the set of all nonnegative integers.
Let $x: \ZZ^{+} \rightarrow \RR^n$.
Then, we often denote $x (t)$, $ t \in  \mathbb{Z^+}$ by a shorthand notation $x$ where no confusion occurs.
$\rho(A)$ denotes the spectral radius, i.e., the maximal magnitude of the eigenvalues of a  real square matrix $A$.
$\bm{1}_N$ denotes an $N$ dimensional column vector whose components are all $1$.
$\otimes$ denotes the Kronecker product of matrices.
$||x||$ denotes the Euclidean norm of a vector $x$
and $\|A\|$ denotes the induced Euclidean norm of a real matrix $A$.
For $X_i\in \RR^{n_i \times q}$, $i=1,\dots,m$,
$
  \text{col}(X_1,\dots,X_m)= \left[
                               \begin{array}{ccc}
                                X_{1}^{T}  & \cdots & X_{m}^{T} \\
                               \end{array}
                             \right]^{T}.
$
%A a real sequence $x(t):\ZZ^{+} \mapsto \RR^{n} $ is said to converge exponentially at a rate faster than $r$ if
%$\|x(t)\| < \alpha r^{t}, t \in \ZZ^{+}$, for some positive numbers $\alpha$, and $0<r<1$.

%%%%%%%%%%%%%%%%%%%%%%%%%%%%%%%%%%%%%%%%%%%%%%%%%%%%%%%%%%%%%%%%%%%%%%%%%%%%%%%%%%%%%%%%%%%%%%%%%%%%%%%%%%%%%%%%%%%%%%%%%%%%%%%%%%%%%%%%%%%%%%%%%%%%%%%%%%%%%%%%%%%%%%%%%%%%%%%%%%%%%%%%%%%%%%%%%%

%%%%%%%%%%%%%%%%%%%%%%%%%%%%%%%%%%%%%%%%%%%%%%%%%%%%%%%%%%%%%%%%%%%%%%%%%%%%%%%%%%%%%%%%%%%%%%%%%%%%%%%%%%%%%%%%%%%%%%%%%%%%%%%%%%%%%%%%%%%%%%%%%%%%%%%%%%%%%%%%%%%%%%%%%%%%%%%%%%%%%%%%%%%%%%%%%%

\section{Some Technical Lemmas}\label{Section Two}

In what follows, we use
$\bar{\mathcal{G}}_{\sigma(t)}=\left(\bar{\mathcal{V}},\bar{\mathcal{E}}_{\sigma(t)}\right)$ to denote a switching digraph\footnote{See Appendix for a summary of notation on digraph.}
with $\bar{\mathcal{V}}=\{0,1,\ldots,N\}$ and
$\sigma: \ZZ^{+} \mapsto \mathcal{P}=\{1,2,\dots,n_{0}\}$.
This digraph is said to be \emph{jointly connected} if there exists
$T \ge 0$ such that, for all $t \in \ZZ^{+}$,
every node $i, i=1,\ldots,N$,
is reachable from the node $0$
in the union digraph $\bigcup_{s=0}^{T} \bar{\mathcal{G}}_{\sigma(t+s)} $.

Let us first list the following two assumptions.

\begin{Assumption}\label{ass-jointly-connected}
The digraph $\bar{\mathcal{G}}_{\sigma(t)}$ is jointly connected.
\end{Assumption}

\begin{Assumption}\label{ass-modolus-le-1}
$ \rho (S) \leq 1$.
\end{Assumption}

Denote the weighted adjacency matrix of $\bar{\mathcal{G}}_{\sigma(t)}$ by
$\bar{\mathcal{A}}_{\sigma(t)}=[a_{ij}(t)]_{i,j=0}^{N}\in \RR^{(N+1) \times (N+1)}$.
Since $\mathcal{P}$ only contains finitely many elements,  there exist real numbers $e_{\max} \ge e_{\min}>0$
such that $e_{\min} \le a_{ij}(t) \le e_{\max}$ for all $t \in \ZZ^{+}$
and $(j,i) \in \bar{\mathcal{E}}_{\sigma(t)}$.

For $i,j = 0, 1, \ldots, N$, let
\begin{equation*}
  \omega_{ij}(t)=
  \begin{cases}
  \frac{1}{1+\sum_{j=0}^{N}a_{ij}(t)},  \quad & \text{if}\  i=j\\
  \frac{a_{ij}(t)}{1+\sum_{j=0}^{N}a_{ij}(t)},  \quad & \text{otherwise.}
  \end{cases}
\end{equation*}
Then, we call
$\bar{\Omega}_{\sigma(t)} = [\omega_{ij}(t)]_{i,j=0}^{N}\in \RR^{(N+1) \times (N+1)} $ as
the \emph{normalized weighted adjacency matrix} of $\bar{\mathcal{G}}_{\sigma(t)}$.

Now consider the following linear switched system:
\begin{equation}\label{eq-x-switched-in-Moreau}
  x(t+1)=\bar{\Omega}_{\sigma(t)}x(t), \qquad    t\ge t_{0} \ge 0
\end{equation}
where $x=\text{col}(x_{0}, x_{1}, \ldots, x_{N})$,
$x_{i}\in \RR, i=0,1,\ldots,N$.

The following proposition is extracted from Proposition 1 in \cite{Moreau05}.

\begin{Proposition}\label{Lemma-Moreau}
Under Assumption \ref{ass-jointly-connected},
for any $x (t_0)$,  the $(N+1)$ components
of the solution $x(t)$ of system (\ref{eq-x-switched-in-Moreau})
converge uniformly to a common value as $t \to \infty$. {\hfill $\Box$\par}
\end{Proposition}

\begin{Remark} \label{Rem1}
Let $\hat{\mathcal{G}}_{\sigma(t)}$ be a subgraph of $\bar{\mathcal{G}}_{\sigma(t)}$,
which is obtained from $\bar{\mathcal{G}}_{\sigma(t)}$
by removing all edges $(i,0)$ for all $i=1,\ldots,N$. Then, Assumption \ref{ass-jointly-connected}
implies that $\hat{\mathcal{G}}_{\sigma(t)}$ is also jointly connected.
\end{Remark}

Now let
$\Lambda_{\sigma(t)} \in \RR^{N \times N}$ consist of the last $N$ rows and the last $N$ columns of
$\bar{\Omega}_{\sigma(t)}$.
Then, we can obtain the following result.

\begin{Lemma} \label{Corollary-exponential-stability}
Under Assumption \ref{ass-jointly-connected},
the linear switched system
\begin{equation}\label{eq-x-switched-reduced}
  y(t+1)=\Lambda_{\sigma(t)} y(t), \qquad   t \ge t_{0} \ge 0
\end{equation}
is exponentially stable.
\end{Lemma}

\begin{Proof}
Let $\hat{\Omega}_{\sigma(t)} \in \RR^{(N+1) \times (N+1)}$ be the normalized weighted adjacency matrix
of $\hat{\mathcal{G}}_{\sigma(t)}$ and
consider the linear switched system
\begin{equation}\label{eq-x-switched-rooted-at-0}
  x(t+1)=\hat{\Omega}_{\sigma(t)}x(t), \qquad    t \ge t_{0} \ge 0
\end{equation}
where $x=\text{col}(x_{0}, x_{1}, \ldots, x_{N})$,
$x_{i}\in \RR, i=0,1,\ldots,N$.
By Remark \ref{Rem1}, $\hat{\mathcal{G}}_{\sigma(t)}$ is also jointly connected.
Thus, by Proposition \ref{Lemma-Moreau},
all components of any solution $x(t)$ of system (\ref{eq-x-switched-rooted-at-0})
converge uniformly to a common value as $t \to \infty$.

Since $\hat{\mathcal{G}}_{\sigma(t)}$ does not contain  such edges as $(i,0),i=1,\ldots,N$,
at any time $t$,   $\hat{\Omega}_{\sigma(t)}$ takes the following form:
\begin{equation*}
  \hat{\Omega}_{\sigma(t)}=\left[
                             \begin{array}{c|c}
                               1 & \mathbf{0}_{1\times N} \\ \hline
                               \Delta_{\sigma(t)} \bm{1}_{N} & \Lambda_{\sigma(t)}  \\
                             \end{array}
                           \right]
\end{equation*}
where $ \Delta_{\sigma(t)}=\text{diag}\{\omega_{10}(t), \ldots \omega_{N0}(t)\}$.
Therefore, $x_{0}(t)=x_{0}(t_{0})$ for all $t\ge t_{0}$.
Thus, $x_{i}(t) \to x_{0}(t_{0}), i=1,\ldots,N$,
uniformly as $t \to \infty$. Letting $x_{0}(t_{0})=0$ shows that all components of
any solution $x(t)$ of system (\ref{eq-x-switched-rooted-at-0})
converge uniformly to the origin.

Now, for any $y (t_0) \in \RR^N$, let $x (t_0) = \text{col} (0, y(t_{0})) \in \RR^{N+1}$.
Then, the last $N$ components of the solution of (\ref{eq-x-switched-rooted-at-0}) starting from $x (t_0)$
coincide with the solution of (\ref{eq-x-switched-reduced}) starting from $y (t_0)$.
Thus, system (\ref{eq-x-switched-reduced}) is uniformly asymptotically stable, or, what is the same, exponentially stable.
\end{Proof}

%%%%%%%%%%%%%%%%%%%%%%%%%%%%%%%%%%%%%%%%%%%%%%%%%%%%%%%%%%%%%%%%%%%%%%%%%%%%%%%%%%%%%%%%%%%%%%%%%%%%%%%%%%%%%%%%%%%%%%%%%%%%%%%%%%%%%%%%%%%%%%%%%%%%%%%%%%%%%%%%%%%%%%%%%%%%%%%%%%%%%%%%%%%%%%%%%%

\begin{Remark}
This lemma can be viewed as a discrete-time counterpart of Corollary 4 in \cite{SuHuangCyber12},
which plays a key role in dealing with the leader-following control problems
for continuous-time multi-agent systems.
It can also be viewed as an extension of Lemma 3.1 in \cite{CCC17} from
connected static networks to jointly connected switching networks.
We believe that this lemma will also play a key role in dealing with the leader-following control problems
for discrete-time multi-agent systems.
\end{Remark}

\begin{Lemma}\label{Lemma-E-S-exponential}
Suppose the following system:
\begin{equation}\label{eq-zeta-E}
  \xi(t+1)=E(t)  \xi(t), \quad    t  \in \ZZ^{+}
\end{equation}
where $E(t)$ is bounded over $\ZZ^{+}$, is exponentially stable .
Then, under Assumption \ref{ass-modolus-le-1},
the following  system:
\begin{equation}\label{eq-zeta-E-S}
  \zeta(t+1)=(E(t) \otimes S) \zeta(t), \quad    t \ge t_{0} \ge 0
\end{equation}
is also exponentially stable.
\end{Lemma}

\begin{Proof}
Given any initial condition $\zeta(t_{0})$,
the solution of system (\ref{eq-zeta-E-S}) can be expressed as
\begin{align}\label{eq-zeta-solution}
  \zeta(t) & = \left( \prod_{s=1}^{t-t_{0}} \left(E(t-s) \otimes S \right) \right) \zeta(t_{0})   \notag  \\
         & = \left( \left(\prod_{s=1}^{t-t_{0}} E(t-s) \right) \otimes S^{t-t_{0}}  \right)  \zeta(t_{0})   \notag  \\
         & = \left( \Phi(t,t_{0})\otimes S^{t-t_{0}}  \right)  \zeta(t_{0}), \qquad t\ge t_{0}
\end{align}
where $\Phi(t,t_{0})$ is the state transition matrix of system (\ref{eq-zeta-E}).

By our assumption on system (\ref{eq-zeta-E}),
there exist positive constants $\alpha_{1}$ and $0 < r_{1} <1$, such that
$
  \|\Phi(t,t_{0})\| \le \alpha_{1} r_{1}^{t-t_{0}}, t\ge t_{0}.
$
Under Assumption \ref{ass-modolus-le-1},
there exists $\epsilon > 0$ such that
$(\rho(S)+\epsilon) r_{1} <1$
and $\left \|S^{t-t_{0}} \right \| < \alpha (\epsilon) \left( \rho (S)+ \epsilon \right)^{t-t_{0}} $ for some real constant
$\alpha (\epsilon)$ \cite{Horn}.
Thus, from (\ref{eq-zeta-solution})
\begin{align*}\label{}
  \|\zeta(t)\| & \le   \left\|\left( \Phi(t,t_{0})\otimes S^{t-t_{0}}  \right) \right\|   \|\zeta(t_{0})\|    \notag \\
             & =   \left\| \Phi(t,t_{0}) \right\|  \left\|S^{t-t_{0}} \right\| \ \|\zeta(t_{0})\|   \notag  \\
             & <  \alpha_{1} \alpha (\epsilon)   \left(\rho (S)+ \epsilon \right)^{t-t_{0}}  r_{1}^{t-t_{0}}  \   \|\zeta(t_{0})\|   \notag  \\
             & = \alpha_{2}  r_{2}^{t-t_{0}}  \   \|\zeta(t_{0})\|, \qquad t \ge t_{0}
\end{align*}
where $\alpha_{2} = \alpha_{1} \alpha (\epsilon) $ and $r_{2} =  (\rho (S)+ \epsilon) r_{1} <1$.
Therefore, system (\ref{eq-zeta-E-S}) is also exponentially stable.
\end{Proof}

\section{Discrete-Time Distributed Observers}\label{Section Three}
In this section, we  study two types of discrete-time distributed observers for the leader system (\ref{eq-leader-system-discrete})
over jointly connected switching networks.

\subsection{Distributed Observer}
Given the switching digraph
$\bar{\mathcal{G}}_{\sigma(t)}=\left(\bar{\mathcal{V}},\bar{\mathcal{E}}_{\sigma(t)}\right)$ and its normalized weighted adjacency matrix
$\bar{\Omega}_{\sigma(t)} = [\omega_{ij}(t)]_{i,j=0}^{N}\in \RR^{(N+1) \times (N+1)}$,
consider the following dynamic compensators:
\begin{equation} \label{eq-eta-distributed-observer}
   \eta_{i}(t+1)  =S\eta_{i}(t) + S \sum_{j =0 }^{N}\omega_{ij}(t)\left(\eta_{j}(t)-\eta_{i}(t)\right), i=1,\ldots,N
\end{equation}
where $\eta_{0}=v$, and, for $i=1,\ldots,N$, $\eta_{i}\in \RR^{q}$.

\begin{Remark}
If, for any initial conditions $v(0)$ and $\eta_{i}(0), i=1,\ldots,N$,
the solutions of systems (\ref{eq-leader-system-discrete}) and (\ref{eq-eta-distributed-observer})
satisfy $\lim_{t \to \infty}(\eta_{i}(t)-v(t))=0, i=1,\ldots,N$,
then system (\ref{eq-eta-distributed-observer})
is called a distributed observer for the leader system (\ref{eq-leader-system-discrete}).
\end{Remark}

\begin{Theorem}\label{Theorem-DO}
Under Assumptions \ref{ass-jointly-connected} and \ref{ass-modolus-le-1},
for any initial conditions $v(0)$ and $\eta_{i}(0), i=1,\ldots,N$,
the solutions of systems (\ref{eq-leader-system-discrete}) and (\ref{eq-eta-distributed-observer})
 satisfy
\begin{equation*}
  \lim_{t \to \infty}(\eta_{i}(t)-v(t))=0, \quad i=1,\ldots,N
\end{equation*}
exponentially.
\end{Theorem}

\begin{Proof}
Let $\tilde{\eta}_{i}=\eta_{i}-v, i=1,\ldots,N$,
and $\tilde{\eta} = \text{col}(\tilde{\eta}_{1}, \ldots, \tilde{\eta}_{N})$.
Then, it can de derived that
\begin{equation}\label{eq-tilde-eta-DO-compact}
  \tilde{\eta}(t+1)=(\Lambda_{\sigma(t)}  \otimes S) \tilde{\eta}(t), \quad t\in \ZZ^{+}.
\end{equation}
By Lemmas \ref{Corollary-exponential-stability} and \ref{Lemma-E-S-exponential},
system (\ref{eq-tilde-eta-DO-compact}) is exponentially stable.
Thus, we have  $\lim_{t \to \infty}(\eta_{i}(t)-v(t))=0, i=1,\ldots,N$,
exponentially.
\end{Proof}

\begin{Remark}\label{Remark-advantages}
The discrete-time distributed observer over jointly connected switching networks
was first studied in \cite{YanHuang17}.
In comparison with the one in \cite{YanHuang17},
the distributed observer (\ref{eq-eta-distributed-observer}) offers at least three advantages.
First, we don't require that the subgraph $\mathcal{G}_{\sigma(t)}=(\mathcal{V},\mathcal{E}_{\sigma(t)})$
of $\bar{\mathcal{G}}_{\sigma(t)}$, where $\mathcal{V}=\{1,\ldots,N\}$
and $\mathcal{E}_{\sigma(t)}=\mathcal{\bar{E}}_{\sigma(t)} \cap (\mathcal{V} \times \mathcal{V})$,
 be undirected for all $t\in \ZZ^{+}$ as in \cite{YanHuang17}.
Second, while the system matrix $S$ in \cite{YanHuang17} is assumed to be marginally stable, we only require $\rho (S) \leq 1$.
Third, while the estimation errors were shown to be asymptotically decaying in \cite{YanHuang17},
we show that the estimation errors decay to zero exponentially.

\end{Remark}

\begin{Remark} It is interesting to note that Theorem \ref{Theorem-DO} implies that, under Assumptions \ref{ass-jointly-connected} and \ref{ass-modolus-le-1},
the leader-following consensus problem with $x_0 (t+1) = S x_0 (t)$ as the leader system and
the following system:
\begin{equation*}
{x}_i (t+1) =  S x_i (t) + u_i (t),\quad  i = 1, \ldots, N \\
\end{equation*}
as $N$ follower subsystems is solvable by the following distributed control law:
\begin{equation*}
   u_i (t)  = S \sum_{j =0 }^{N}\omega_{ij}(t)\left(x_{j}(t)-x_{i}(t)\right),\  i=1,\ldots,N.
\end{equation*}
This problem has been an open problem until now.
In particular, it includes the leader-following consensus problem of multiple
discrete-time double-integrator systems as a special case.
It is also interesting to note that the leaderless consensus problem for linear multi-agent systems over jointly switching networks
was studied in \cite{QinGao14} and \cite{Qin15}.
However, since our result relies on our newly established key lemma (Lemma \ref{Corollary-exponential-stability}), the approach in \cite{QinGao14} and \cite{Qin15} cannot lead to
Theorem \ref{Theorem-DO} directly.
\end{Remark}

\subsection{Adaptive Distributed Observer}
The distributed observer (\ref{eq-eta-distributed-observer}) assumes that the control $u_i$ of every follower subsystem knows the system matrix $S$
of the leader system. In practice, such information may not be available for all follower subsystems
for all $t\in \ZZ^{+}$.
Therefore, we further propose the following so-called
adaptive distributed observer candidate:
\begin{subequations}\label{eq-adaptive-distributed-observer}
\begin{align}
  S_{i}(t+1) & = S_{i}(t) +  \sum_{j =0}^{N}\omega_{ij}(t)\left(S_{j}(t)-S_{i}(t)\right), \ i=1,\ldots,N         \label{eq-matrix-estimation}  \\
  \eta_{i}(t+1) & =S_{i}(t)\eta_{i}(t) + S_{i}(t)\sum_{j =0 }^{N}\omega_{ij}(t)\left(\eta_{j}(t)-\eta_{i}(t)\right)  \label{eq-state-estimation}
\end{align}
\end{subequations}
where $ S_{0}=S, \eta_{0}=v $, and, for $i=1,\ldots,N$, $S_{i} \in \RR^{q \times q}, \eta_{i} \in \RR^{q}$.

%%%%%%%%%%%%%%%%%%%%%%%%%%%%%%%%%%%%%%%%%%%%%%%%%%%%%%%%%%%%%%%%%%%%%%%%%%%%%%%%%%%%%%%%%%%%%%%%%%%%%%%%%%%%%%%%%%%%%%%%%%%%%%%%%%%%%%%%%%%%%%%%%%%%%%%%%%%%%%%%%%%%%%%%%%%%%%%%%%%%%%%%%%%%%%%%%%
\begin{Remark}
If, for any initial conditions $v(0)$ and $S_{i}(0)$, $\eta_{i}(0), i=1,\ldots,N$,
the solutions of systems (\ref{eq-leader-system-discrete}) and (\ref{eq-adaptive-distributed-observer})
satisfy $\lim_{t \to \infty}(S_{i}(t)-S)=0$ and $\lim_{t \to \infty}(\eta_{i}(t)-v(t))=0, i=1,\ldots,N$,
then system (\ref{eq-adaptive-distributed-observer})
is called an adaptive distributed observer for the leader system (\ref{eq-leader-system-discrete}).
It can be seen from (\ref{eq-matrix-estimation}) that only those
followers with $\omega_{i0}(t) \ne 0, i=1,\ldots,N$, know the system matrix $S$ of the leader system at
the time instant $t$.
\end{Remark}

%%%%%%%%%%%%%%%%%%%%%%%%%%%%%%%%%%%%%%%%%%%%%%%%%%%%%%%%%%%%%%%%%%%%%%%%%%%%%%%%%%%%%%%%%%%%%%%%%%%%%%%%%%%%%%%%%%%%%%%%%%%%%%%%%%%%%%%%%%%%%%%%%%%%%%%%%%%%%%%%%%%%%%%%%%%%%%%%%%%%%%%%%%%%%%%%%%%%%%%%%%%%%%%%%%%%%

Before we state the next theorem, we quote  Lemma 1 in  \cite{LTHuang17}
as follows.

\begin{Lemma}\label{Lemma-outside-perturbed-linear-system-preliminary}
Consider the following system:
\begin{equation}\label{eq-outside-perturbed-linear-system-preliminary}
  z(t+1)=C(t) z(t) + d(t), \qquad   t \ge t_{0} \ge 0
\end{equation}
where $C(t)$ and $d(t)$ are bounded over $\ZZ^{+}$.
Suppose the nominal system
\begin{equation*}\label{eq-nominal-system-on-z}
  z(t+1)=C(t) z(t),  \qquad t\in \ZZ^{+}
\end{equation*}
is exponentially stable
and $d(t) \to 0$ exponentially as $t \to \infty$.
Then, for any initial condition $ z(t_{0})$, the solution $z(t)$ of system (\ref{eq-outside-perturbed-linear-system-preliminary})
converges to
the origin exponentially.
{\hfill $\Box$\par}
\end{Lemma}

%%%%%%%%%%%%%%%%%%%%%%%%%%%%%%%%%%%%%%%%%%%%%%%%%%%%%%%%%%%%%%%%%%%%%%%%%%%%%%%%%%%%%%%%%%%%%%%%%%%%%%%%%%%%%%%%%%%%%%%%%%%%%%%%%%%%%%%%%%%%%%%%%%%%%%%%%%%%%%%%%%%%%%%%%%%%%%%%%%%%%%%%%%%%%%%%%%%%%%%%%%%%%%%%%%%%%%%

\begin{Theorem}\label{Thrm-Adaptive-distributed-Observer}
Consider systems (\ref{eq-leader-system-discrete}) and (\ref{eq-adaptive-distributed-observer}).
For any initial conditions $v(0)$ and $S_{i}(0)$, $\eta_{i}(0), i=1,\ldots,N$
\begin{enumerate}
\item[(i)] under Assumption \ref{ass-jointly-connected},
the solution of system (\ref{eq-matrix-estimation})  satisfies
\begin{equation*}\label{}
  \lim_{t\to \infty} (S_{i}(t)-S)=0, \quad i=1,\ldots,N
\end{equation*}
exponentially;
\item[(ii)] under Assumptions  \ref{ass-jointly-connected} and \ref{ass-modolus-le-1},
the solutions of systems (\ref{eq-leader-system-discrete})  and  (\ref{eq-state-estimation}) satisfy
\begin{equation*}\label{}
  \lim_{t \to \infty} (\eta_{i}(t)-v(t)) = 0, \quad i=1,\ldots,N
\end{equation*}
exponentially.
\end{enumerate}
\end{Theorem}

%%%%%%%%%%%%%%%%%%%%%%%%%%%%%%%%%%%%%%%%%%%%%%%%%%%%%%%%%%%%%%%%%%%%%%%%%%%%%%%%%%%%%%%%%%%%%%%%%%%%%%%%%%%%%%%%%%%%%%%%%%%%%%%%%%%%%%%%%%%%%%%%%%%%%%%%%%%%%%%%%%%%%%%%%%%%%%%%%%%%%%%%%%%%%%%%%%%%%%%%%%%%%%%%%%%%%%%

\begin{Proof}
\emph{Part (i).}
Let $\tilde{S}_{i}=S_{i}-S, i=1,\ldots,N$,
and $\tilde{S}=\text{col}(\tilde{S}_{1},\ldots, \tilde{S}_{N})$.
Then, system (\ref{eq-matrix-estimation}) can be put into the following compact form:
\begin{equation}\label{eq-S-compact}
  \tilde{S}(t+1)=(\Lambda_{\sigma(t)}   \otimes I_{q})  \tilde{S}(t), \quad t \in \ZZ^{+}.
\end{equation}
By Lemma \ref{Corollary-exponential-stability}, system (\ref{eq-S-compact})
is exponentially stable.
Thus, we have
$
  \lim_{t\to \infty} (S_{i}(t)-S)=0,  i=1,\ldots,N,
$
exponentially.

%%%%%%%%%%%%%%%%%%%%%%%%%%%%%%%%%%%%%%%%%%%%%%%%%%%%%%%%%%%%%%%%%%%%%%%%%%%%%%%%%%%%%%%%%%%%%%%%%%%%%%%%%%%%%%%%%%%%%%%%%%%%%%%%%%%%%%%%%%%%%%%%%%%%%%%%%%%%%%%%%%%%%%%%%%%%%%%%%%%%%%%%%%%%%%%%%%%%%%%%%%%%%%%%%%%%%%%

\emph{Part (ii)}. For $i=1,\ldots,N$, let  $\tilde{\eta}_{i}=\eta_{i}-v$.
Then, it can be derived from (\ref{eq-state-estimation}) that
\begin{align}\label{eq-astimation-error-of-state}
  \tilde{\eta}_{i}(t+1)
   & = S \tilde{\eta}_{i} + \tilde{S}_{i}\eta_{i}+ S_{i}\sum_{j =0}^{N} \omega_{ij}(t)(\tilde{\eta}_{j}-\tilde{\eta}_{i})   \notag  \\
   & = S \tilde{\eta}_{i} + S\sum_{j=0}^{N} \omega_{ij}(t)(\tilde{\eta}_{j}-\tilde{\eta}_{i})+ \tilde{S}_{i}v   \notag \\
   & \quad + \tilde{S}_{i} \tilde{\eta}_{i} + \tilde{S}_{i}\sum_{j  =0}^{N} \omega_{ij}(t)(\tilde{\eta}_{j}-\tilde{\eta}_{i}).
\end{align}
Let $\tilde{\eta} = \text{col}(\tilde{\eta}_{1}, \ldots, \tilde{\eta}_{N})$ and $\tilde{S}_{d} = \text{block diag} \{\tilde{S}_{1}, \ldots, \tilde{S}_{N}\}$.
Then, system (\ref{eq-astimation-error-of-state}) can be put into the following compact form:
\begin{align}\label{eq-estimation-error-of-state-compact}
  \tilde{\eta}(t+1) & =  \left(\Lambda_{\sigma(t)}  \otimes S  \right  ) \tilde{\eta} + \tilde{S}_{d} (\bm{1}_{N} \otimes v)    \notag  \\
   & \quad + \left(\tilde{S}_{d} +  \left[
                                       \begin{array}{c}
                                         (\Lambda_{\sigma(t)}-I_{N})_{1} \otimes \tilde{S}_{1} \\
                                         \vdots \\
                                         (\Lambda_{\sigma(t)}-I_{N})_{N} \otimes \tilde{S}_{N} \\
                                       \end{array}
                                     \right]
      \right)\tilde{\eta}
\end{align}
where $ (\Lambda_{\sigma(t)}-I_{N})_{i}, i=1,\ldots,N$,
denotes the $i$th row of the matrix $(\Lambda_{\sigma(t)}-I_{N})$.

Let
\begin{align*}
  \Gamma_{1}(t) & =  \Lambda_{\sigma(t)}  \otimes S, \quad  \Gamma_{3}(t)  = \tilde{S}_{d}(t) (\bm{1}_{N} \otimes v(t))  \\
  \Gamma_{2}(t) & =  \tilde{S}_{d}(t) +  \left[
                                       \begin{array}{c}
                                         ( \Lambda_{\sigma(t)}-I_{N})_{1} \otimes \tilde{S}_{1}(t) \\
                                         \vdots \\
                                         (\Lambda_{\sigma(t)}-I_{N})_{N} \otimes \tilde{S}_{N}(t) \\
                                       \end{array}
                                     \right] .
\end{align*}
Then, system (\ref{eq-estimation-error-of-state-compact}) becomes
\begin{equation}\label{eq-estimation-error-of-state-succinct}
  \tilde{\eta}(t+1)=\left( \Gamma_{1}(t)+\Gamma_{2}(t)  \right )\tilde{\eta}(t) + \Gamma_{3}(t),
  \quad t \in \ZZ^{+}.
\end{equation}

By Part (i), $\lim_{t\to\infty}\tilde{S}_{d} (t)=0$ exponentially.
Thus,  $\lim_{t\to\infty} \Gamma_2 (t)=0$ exponentially, and,  under Assumption \ref{ass-modolus-le-1}, $\lim_{t \to\infty} \Gamma_3 (t) =0$ exponentially.
As shown in Theorem \ref{Theorem-DO}, the system
$
  \tilde{\eta}(t+1)= \Gamma_{1}(t)\tilde{\eta}(t)
$
is exponentially stable. Since $\lim_{t\to\infty} \Gamma_2 (t)=0$ exponentially,
by Theorem 24.7 of \cite{Rugh96},
  the  system $
  \tilde{\eta}(t+1)= \left(\Gamma_{1}(t) + \Gamma_{2}(t) \right)\tilde{\eta}(t)$ is also exponentially stable.
Since $\lim_{t\to\infty} \Gamma_3 (t)=0$ exponentially, it follows from
Lemma \ref{Lemma-outside-perturbed-linear-system-preliminary} that  the solution of the system (\ref{eq-estimation-error-of-state-succinct})
converges to the origin exponentially.
Thus, we have
$
  \lim_{t\to \infty} (\eta_{i}(t)-v(t))=0,  i=1,\ldots,N,
$
exponentially.
\end{Proof}

%%%%%%%%%%%%%%%%%%%%%%%%%%%%%%%%%%%%%%%%%%%%%%%%%%%%%%%%%%%%%%%%%%%%%%%%%%%%%%%%%%%%%%%%%%%%%%%%%%%%%%%%%%%%%%%%%%%%%%%%%%%%%%%%%%%%%%%%%%%%%%%%%%%%%%%%%%%%%%%%%%%%%%%%%%%%%%%%%%%%%%%%%%%%%%%%%%%%%%%%%%%%%%%%%%%%%%%

\begin{Remark}
As a result of Theorem \ref{Thrm-Adaptive-distributed-Observer}, we call system (\ref{eq-adaptive-distributed-observer})
an adaptive distributed observer for the leader system (\ref{eq-leader-system-discrete}).
For the special case where the digraph $\bar{\mathcal{G}}_{\sigma(t)}$ is static,
the adaptive distributed observer (\ref{eq-adaptive-distributed-observer})
reduces to the one recently developed in \cite{LTHuang19}.
\end{Remark}

%%%%%%%%%%%%%%%%%%%%%%%%%%%%%%%%%%%%%%%%%%%%%%%%%%%%%%%%%%%%%%%%%%%%%%%%%%%%%%%%%%%%%%%%%%%%%%%%%%%%%%%%%%%%%%%%%%%%%%%%%%%%%%%%%%%%%%%%%%%%%%%%%%%%%%%%%%%%%%%%%%%%%%%%%%%%%%%%%%%%%%%%%%%%

\section{An Application}\label{Section Four}
The cooperative output regulation problem is
an extension of the classical output regulation problem
\cite{Davison76}, \cite{Francis77},  \cite{Huang04}
from a single plant to a multi-agent system, and was first formulated and studied in \cite{SuHuang12}.
It can also be viewed as an extension of the leader-following consensus problem as studied in
\cite{You2013}, \cite{HuHong07}, \cite{NiCheng10} in the sense that,
it not only achieves the asymptotical tracking, but also the disturbance rejection,
where both the reference input and the disturbance
are generated by the leader system.

In this section, we apply the distributed observer
developed in Section \ref{Section Three} to solve the  discrete-time
cooperative linear output regulation problem over jointly connected switching networks.
We note that it is also possible to apply the adaptive distributed observer approach
to solve the  discrete-time
cooperative linear output regulation problem over jointly connected switching networks by referring to
\cite{LTHuang17}.
%%%%%%%%%%%%%%%%%%%%%%%%%%%%%%%%%%%%%%%%%%%%%%%%%%%%%%%%%%%%%%%%%%%%%%%%%%%%%%%%%%%%%%%%%%%%%%%%%%%%%%%%%%%%%%%%%%%%%%%%%%%%%%%%%%%%%%%%%%%%%%%%%%%%%%%%%%%%%%%%%%%%%%%%%%%%%%%%%%%%%%%%%%%%

\subsection{Problem Formulation}
Consider the following discrete-time linear system:
\begin{align}\label{eq-certain-followers}
  x_{i}(t+1) & =A_{i}x_{i}(t) + B_{i}u_{i}(t) +E_{i}v(t), \quad  i=1,\ldots,N  \notag   \\
  e_{i}(t) & = C_{i}x_{i}(t) + D_{i}u_{i}(t) + F_{i}v(t),  \quad   t  \in \ZZ^{+}
\end{align}
where,
for $ i=1,\ldots,N$,  $x_{i} \in \RR^{n_{i}}, u_{i} \in \RR^{m_{i}}$, and
$e_{i} \in \RR^{p_{i}}$ are the state, control input, and
regulated output of the $i$th subsystem, respectively;
$v \in \RR^{q}$ is the state of the exosystem (\ref{eq-leader-system-discrete}) representing the reference input to be tracked and/or the external disturbance to be rejected;
matrices $A_{i}, B_{i}, C_{i}, D_{i}, E_{i},$ and
$F_{i}$ are constant with compatible dimensions.

%%%%%%%%%%%%%%%%%%%%%%%%%%%%%%%%%%%%%%%%%%%%%%%%%%%%%%%%%%%%%%%%%%%%%%%%%%%%%%%%%%%%%%%%%%%%%%%%%%%%%%%%%%%%%%%%%%%%%%%%%%%%%%%%%%%%%%%%%%%%%%%%%%%%%%%%%%%%%%%%%%%%%%%%%%%%%%%%%%%%%%%%%%%%

Like in \cite{YanHuang17}, we treat the system composed of (\ref{eq-leader-system-discrete}) and (\ref{eq-certain-followers}) as a multi-agent system of $(N+1)$ agents with system (\ref{eq-leader-system-discrete}) as the leader and the $N$ subsystems of (\ref{eq-certain-followers}) as $N$ followers.
The network topology of this multi-agent system is described by a switching digraph $\bar{\mathcal{G}}_{\sigma(t)}=\left(\bar{\mathcal{V}},\bar{\mathcal{E}}_{\sigma(t)}\right)$
where $\bar{\mathcal{V}}=\{0,1,\ldots,N\}$ with
the node $0$ associated with the leader system (\ref{eq-leader-system-discrete})
and the node $i,i=1,\ldots,N,$ associated with the $i$th follower subsystem of (\ref{eq-certain-followers}), and,
for $i,j= 0, 1,\ldots,N,$  $(j,i) \in \bar{\mathcal{E}}_{\sigma(t)}$
if and only if  agent $i$ can use the information of agent $j$ for control
at the time instant $t$.
We consider the following class of so-called distributed control laws:
\begin{align}\label{eq-certain-control-law-general}
  u_{i}(t) & =k_{i}(x_{i}(t),\xi_{i}(t)), \quad i=1,\ldots,N  \notag  \\
  \xi_{i}(t+1) &= g_{i}\left(\xi_{i}(t), \xi_{j}(t), j \in \bar{\mathcal{N}}_{i}(t)\right)
\end{align}
where $\xi_{0}=v$, and,  for $i=1,\ldots,N $,  $k_{i}$, $g_i$ are some linear functions of their arguments
and $\bar{\mathcal{N}}_{i}(t)$ denotes the neighbor set of the node $i$ at the time instant $t$.

Now, we are ready to describe the problem.

%%%%%%%%%%%%%%%%%%%%%%%%%%%%%%%%%%%%%%%%%%%%%%%%%%%%%%%%%%%%%%%%%%%%%%%%%%%%%%%%%%%%%%%%%%%%%%%%%%%%%%%%%%%%%%%%%%%%%%%%%%%%%%%%%%%%%%%%%%%%%%%%%%%%%%%%%%%%%%%%%%%%%%%%%%%%%%

\begin{Problem}\label{Problem}
Given the leader system (\ref{eq-leader-system-discrete}), the follower system (\ref{eq-certain-followers}),
and a switching digraph $\bar{\mathcal{G}}_{\sigma(t)}$,
find a distributed control law of the form (\ref{eq-certain-control-law-general}) such that, for $i=1,\ldots,N$, and
 any $x_{i}(0)$, $\xi_{i}(0)$ and $v(0)$
\begin{enumerate}
\item[1)] the solution of the closed-loop system is bounded over $\ZZ^{+}$ when $v(t)$ is bounded over $\ZZ^{+}$;
\item[2)] the regulated output $e_{i}(t)$ satisfies
$ \lim_{t \to \infty} e_{i}(t) = 0$.
\end{enumerate}
\end{Problem}

%%%%%%%%%%%%%%%%%%%%%%%%%%%%%%%%%%%%%%%%%%%%%%%%%%%%%%%%%%%%%%%%%%%%%%%%%%%%%%%%%%%%%%%%%%%%%%%%%%%%%%%%%%%%%%%%%%%%%%%%%%%%%%%%%%%%%%%%%%%%%%%%%%%%%%%%%%%%%%%%%%%%%%%%%%%%%%%%%%%%%%%%

In addition to Assumptions \ref{ass-jointly-connected} and \ref{ass-modolus-le-1},
we list two more assumptions as follows.

\begin{Assumption}\label{ass-certain-stabilizable}
The pairs $(A_{i}, B_{i}), i=1, \ldots, N$, are stabilizable.
\end{Assumption}

\begin{Assumption}\label{ass-solution-to-regulator-equations}
The linear matrix equations
\begin{align}\label{eq-certain-regulator-equations}
  X_{i}S & = A_{i}X_{i} +B_{i}U_{i} + E_{i}   \notag   \\
  \mathbf{0}  & = C_{i}X_{i} +D_{i}U_{i} + F_{i}
\end{align}
have solution pairs $\left(X_{i}, U_{i}\right)$, $i=1,\ldots,N$.
\end{Assumption}

\begin{Remark}
In the classical output regulation problem \cite{Francis77}, \cite{Huang04},
equations (\ref{eq-certain-regulator-equations}) are called the regulator equations,
whose solvability imposes a necessary condition for the solvability of the output regulation problem.
\end{Remark}

%%%%%%%%%%%%%%%%%%%%%%%%%%%%%%%%%%%%%%%%%%%%%%%%%%%%%%%%%%%%%%%%%%%%%%%%%%%%%%%%%%%%%%%%%%%%%%%%%%%%%%%%%%%%%%%%%%%%%%%%%%%%%%%%%%%%%%%%%%%%%%%%%%%%%%%%%%%%%%%%%%%%%%%%%%%%%%%%%%%%%%%%%%%%%%%%%%%%%%%%%

\subsection{Solvability of the Problem}

%%%%%%%%%%%%%%%%%%%%%%%%%%%%%%%%%%%%%%%%%%%%%%%%%%%%%%%%%%%%%%%%%%%%%%%%%%%%%%%%%%%%%%%%%%%%%%%%%%%%%%%%%%%%%%%%%%%%%%%%%%%%%%%%%%%%%%%%%%%%%%%%%%%%%%%%%%%%%%%%%%%%%%%%%%%%%%%%%%%%%%%%%%%%%%%%%%%%
For $i=1,\ldots,N$, under Assumption \ref{ass-certain-stabilizable},
let $K_{xi}$ be such that $\left(A_{i}+B_{i}K_{xi}\right)$ is Schur.
Further, under Assumption \ref{ass-solution-to-regulator-equations},
let  $K_{v i}$ be given by
\begin{equation}\label{eq-K-gain-relation}
    K_{v i}=U_{i} - K_{xi} X_{i}.
\end{equation}
Then,  we design the following distributed control law:
\begin{subequations}\label{eq-overall-control-law}
\begin{align}
  u_{i}(t) &=K_{xi}x_{i}(t) + K_{v i} \eta_{i}(t), \quad  i=1,\ldots,N   \label{eq-state-feedback-control-law} \\
   \eta_{i}(t+1)  &=S\eta_{i}(t) + S \sum_{j =0 }^{N}\omega_{ij}(t)\left(\eta_{j}(t)-\eta_{i}(t)\right) .
\end{align}
\end{subequations}

%%%%%%%%%%%%%%%%%%%%%%%%%%%%%%%%%%%%%%%%%%%%%%%%%%%%%%%%%%%%%%%%%%%%%%%%%%%%%%%%%%%%%%%%%%%%%%%%%%%%%%%%%%%%%%%%%%%%%%%%%%%%%%%%%%%%%%%%%%%%%%%%%%%%%%%%%%%%%%%%%%%%%%%%%%%%%%%%%%%%%%%%%%%%%%%%%%%

\begin{Theorem}\label{Theo-certain-state-feedback}
Under Assumptions \ref{ass-jointly-connected} to \ref{ass-solution-to-regulator-equations},
Problem \ref{Problem} is solvable by the distributed control law (\ref{eq-overall-control-law}).
\end{Theorem}

%%%%%%%%%%%%%%%%%%%%%%%%%%%%%%%%%%%%%%%%%%%%%%%%%%%%%%%%%%%%%%%%%%%%%%%%%%%%%%%%%%%%%%%%%%%%%%%%%%%%%%%%%%%%%%%%%%%%%%%%%%%%%%%%%%%%%%%%%%%%%%%%%%%%%%%%%%%%%%%%%%%%%%%%%%%%%%%%%%%%%%%%%%%%%%%%%%%%%%

\begin{Proof}
For $i=1,\ldots,N$, let
$ \tilde{x}_{i} = x_{i} - X_{i}v$  and
$ \tilde{u}_{i} =u_{i}-U_{i}v$.
Then, by making use of the solution to the regulator equations (\ref{eq-certain-regulator-equations}),
we obtain
\begin{align}\label{eq-certain-error-state-xi}
  \tilde{x}_{i}(t+1)     & = A_{i}\left(\tilde{x}_{i}(t) + X_{i}v(t) \right) + B_{i}\left(\tilde{u}_{i}(t) + U_{i}v(t)\right)  \notag \\
     & \quad +E_{i}v(t) -X_{i}Sv(t)   \notag \\
     & = A_{i}\tilde{x}_{i}(t) + B_{i}\tilde{u}_{i}(t)
\end{align}
and
\begin{align}\label{eq-certain-error-ei}
  e_{i}(t)   &= C_{i}\left(\tilde{x}_{i}(t) + X_{i}v(t) \right) +D_{i}\left(\tilde{u}_{i}(t) + U_{i}v(t)\right) + F_{i}v(t) \notag \\
           &= C_{i}\tilde{x}_{i}(t)+ D_{i} \tilde{u}_{i}(t).
\end{align}
Next, by (\ref{eq-K-gain-relation}) and (\ref{eq-state-feedback-control-law}), we have
\begin{align}\label{eq-certain-error-control-ui}
 \tilde{u}_{i}(t)    &= K_{xi} \tilde{x}_{i}(t)+ K_{v i}(\eta_{i}(t)-v(t)).
\end{align}
Substituting (\ref{eq-certain-error-control-ui}) into (\ref{eq-certain-error-state-xi}) gives
\begin{equation*}\label{}
  \tilde{x}_{i}(t+1)  = \left(A_{i}+B_{i}K_{xi}\right)\tilde{x}_{i}(t) + B_{i}K_{v i}(\eta_{i}(t)-v(t)).
\end{equation*}
By Theorem \ref{Theorem-DO},
$\lim_{t \to \infty}(\eta_{i}(t)-v(t))=0$ exponentially.
Moreover, since $\left(A_{i}+B_{i}K_{xi}\right)$ is Schur, by Lemma \ref{Lemma-outside-perturbed-linear-system-preliminary},
for any initial condition $\tilde{x}_{i}(0)$, $\lim_{t \to \infty} \tilde{x}_{i}(t)=0$ exponentially.
As a result, $\lim_{t \to \infty} \tilde{u}_{i}(t)=0$ exponentially by (\ref{eq-certain-error-control-ui}),
and hence
$\lim_{t \to \infty} e_{i}(t) =0$
exponentially by (\ref{eq-certain-error-ei}).
\end{Proof}

%%%%%%%%%%%%%%%%%%%%%%%%%%%%%%%%%%%%%%%%%%%%%%%%%%%%%%%%%%%%%%%%%%%%%%%%%%%%%%%%%%%%%%%%%%%%%%%%%%%%%%%%%%%%%%%%%%%%%%%%%%%%%%%%%%%%%%%%%%%%%%%%%%%%%%%%%%%%%%%%%%%%%%%%%%%%%%%%%%%%%%%%%%%%%%%%%%%%%%%%%%%

%%%%%%%%%%%%%%%%%%%%%%%%%%%%%%%%%%%%%%%%%%%%%%%%%%%%%%%%%%%%%%%%%%%%%%%%%%%%%%%%%%%%%%%%%%%%%%%%%%%%%%%%%%%%%%%%%%%%%%%%%%%%%%%%%%%%%%%%%%%%%%%%%%%%%%%%%%%%%%%%%%%%%%%%%%%%%%%%%%%%%%%%%%%%%%%%%%

\section{An Example}\label{Section Five}
\begin{figure}
  \centering
  \includegraphics[scale=0.25]{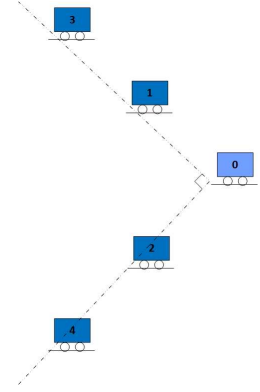}\\
  \caption{The desired leader-following formation of mobile robots}\label{Fig-formation}
\end{figure}

In this section, we consider a leader-following formation problem of five mobile robots.
Let the trajectory of the leader  be generated by the following leader system:
\begin{align*}
    v(t+1)& = Sv(t)=\left(\left[
                       \begin{array}{cc}
                         1 & 1 \\
                         0 & 1 \\
                       \end{array}
                     \right] \otimes I_{2}
  \right) v(t)      \notag  \\
 \left[
    \begin{array}{c}
      x_{0}(t)  \\
       y_{0}(t)\\
    \end{array}
  \right] & = \left( \left[
                     \begin{array}{cc}
                       1 & 0 \\
                     \end{array}
                   \right] \otimes I_{2}
   \right) v(t), \quad t\in \ZZ^{+}
\end{align*}
with the initial condition $v(0)=[x_{d0},y_{d0}, w_{x0},w_{y0}]^{T}$.
The four followers are described by double-integrators:
\begin{align}\label{eq-follower}
  x_{i}(t+1) & = x_{i}(t)+w_{xi}(t) \notag \\
  y_{i}(t+1) & =y_{i}(t)+w_{yi} (t)   \notag\\
  w_{xi}(t+1) & =w_{xi}(t)+ u_{xi}(t) ,  \quad i=1,2,3,4 \notag \\
  w_{yi}(t+1) & =w_{yi}(t) + u_{yi}(t) ,   \quad t \in \ZZ^{+}.
\end{align}
The objective is to design a distributed control law such that
the leader and the four followers  will asymptotically form a geometric pattern as shown in Figure \ref{Fig-formation}, or mathematically,
\begin{align}\label{eq-objective}
 & \lim_{t\to \infty}\left( \left[
                       \begin{array}{c}
                         x_{i}(t) \\
                         y_{i}(t) \\
                       \end{array}
                     \right] -  \left[
                       \begin{array}{c}
                         x_{0}(t) \\
                         y_{0}(t) \\
                       \end{array}
                     \right] \right)   =\left[
                                        \begin{array}{c}
                                          x_{di} \\
                                           y_{di} \\
                                        \end{array}
                                      \right]   \notag \\
  &  \lim_{t\to \infty} \left( \left[
                       \begin{array}{c}
                         w_{xi}(t) \\
                         w_{yi}(t) \\
                       \end{array}
                     \right] -\left[
                       \begin{array}{c}
                         w_{x0} \\
                         w_{y0} \\
                       \end{array}
                     \right] \right)  =  0, \quad i=1,2,3,4
\end{align}
in which $[x_{di}, y_{di}]^{T}, i=1,2,3,4$, denotes the desired
constant relative position between the $i$th follower and the leader.

Define the regulated output of each follower as
\begin{equation}\label{eq-regulated-error}
  e_{i}(t)=\left[
          \begin{array}{c}
             x_{i}(t)-x_{di} \\
             y_{i}(t)-y_{di} \\
          \end{array}
        \right]-\left[
                  \begin{array}{c}
                    x_{0}(t) \\
                    y_{0}(t) \\
                  \end{array}
                \right], \quad i=1,2,3,4.
\end{equation}
Then, the system composed of (\ref{eq-follower}) and (\ref{eq-regulated-error})
is in the form of  (\ref{eq-certain-followers})
with the state $[x_{i}-x_{di},y_{i}-y_{di},w_{xi},w_{yi}]^{T}$,
regulated output $e_{i}$, control input $u_{i}=[u_{xi},u_{yi}]^{T}$,
and various matrices given by
\begin{align*}
  A_{i} & = \left[
              \begin{array}{cc}
                1 & 1 \\
                0 & 1 \\
              \end{array}
            \right] \otimes I_{2},  \quad B_{i}=\left[
                                                  \begin{array}{c}
                                                    0 \\
                                                    1 \\
                                                  \end{array}
                                                \right]\otimes I_{2},
                                                \quad E_{i}=\mathbf{0}_{4 \times 4}
   \\
  C_{i} & =\left[
             \begin{array}{cc}
               1 & 0 \\
             \end{array}
           \right]\otimes I_{2}, \quad D_{i}=\mathbf{0}_{2 \times 2}, \quad
           F_{i}=- \left[
                     \begin{array}{cc}
                       1 & 0 \\
                     \end{array}
                   \right]\otimes I_{2}.
\end{align*}
In fact, the objective in (\ref{eq-objective}) can be achieved if the
corresponding cooperative output regulation problem is solvable.

\begin{figure}
  \centering
  \subfigure[$\bar{\mathcal{G}}_{1}$]{
    \includegraphics[width=0.72in]{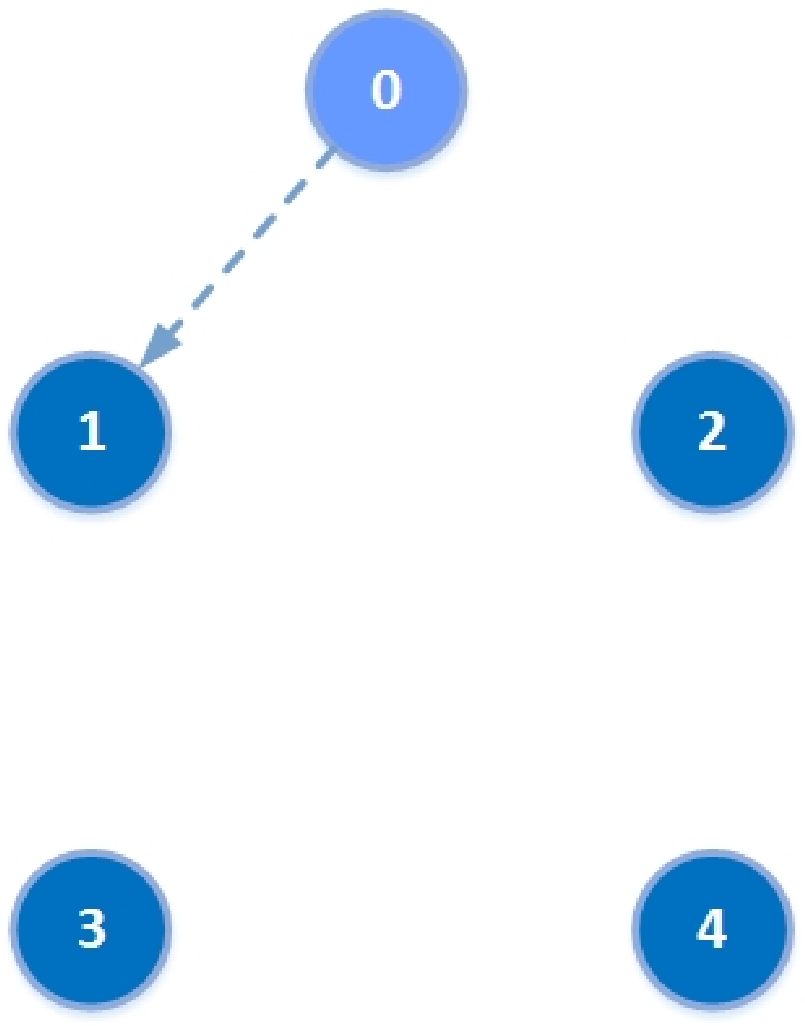}}
  \hfill
  \subfigure[$\bar{\mathcal{G}}_{2}$]{
    \includegraphics[width=0.72in]{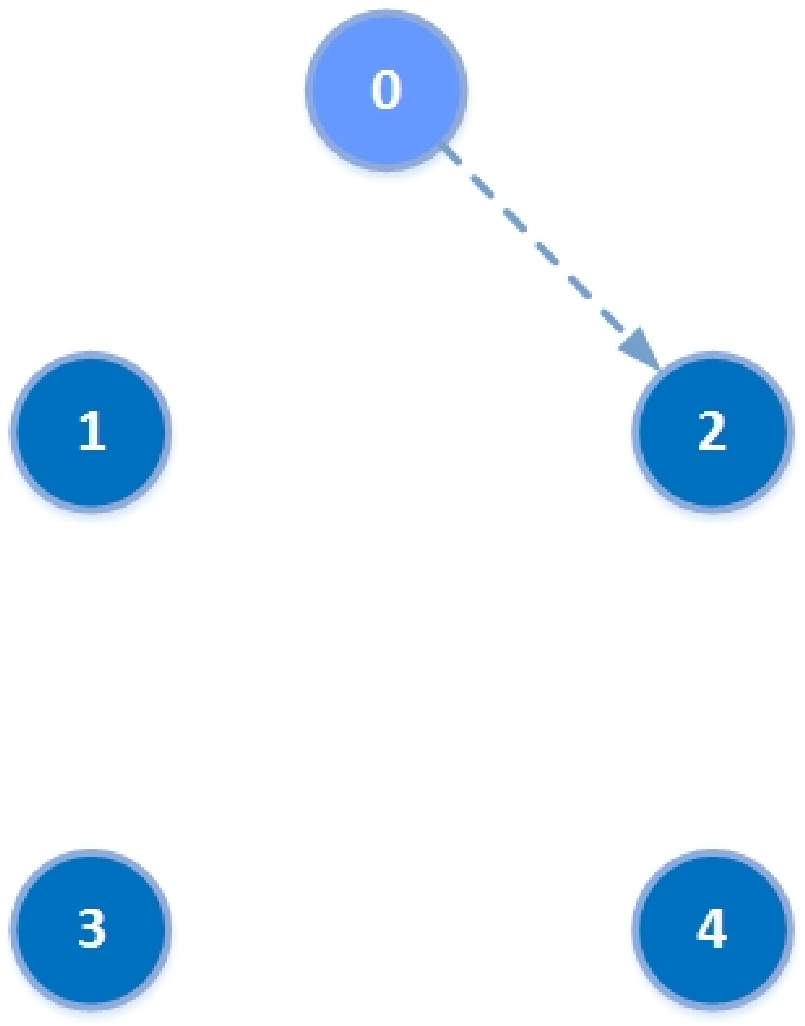}}
\hfill
  \subfigure[$\bar{\mathcal{G}}_{3}$]{
    \includegraphics[width=0.72in]{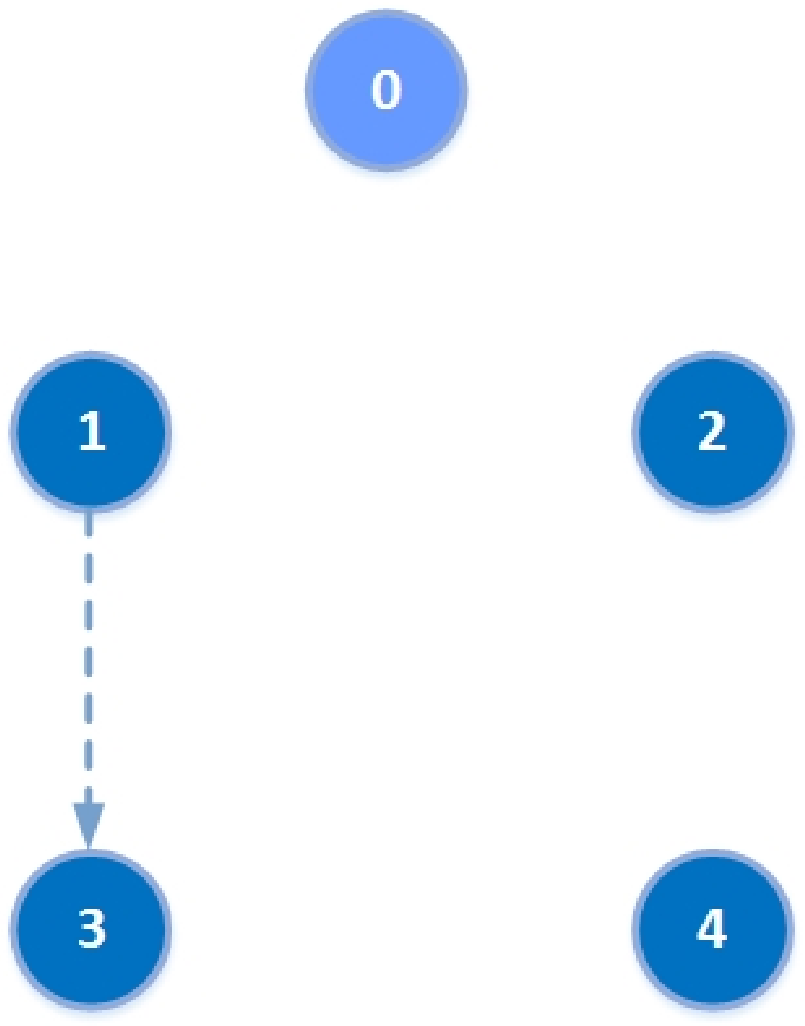}}
\hfill
  \subfigure[$\bar{\mathcal{G}}_{4}$]{
    \includegraphics[width=0.72in]{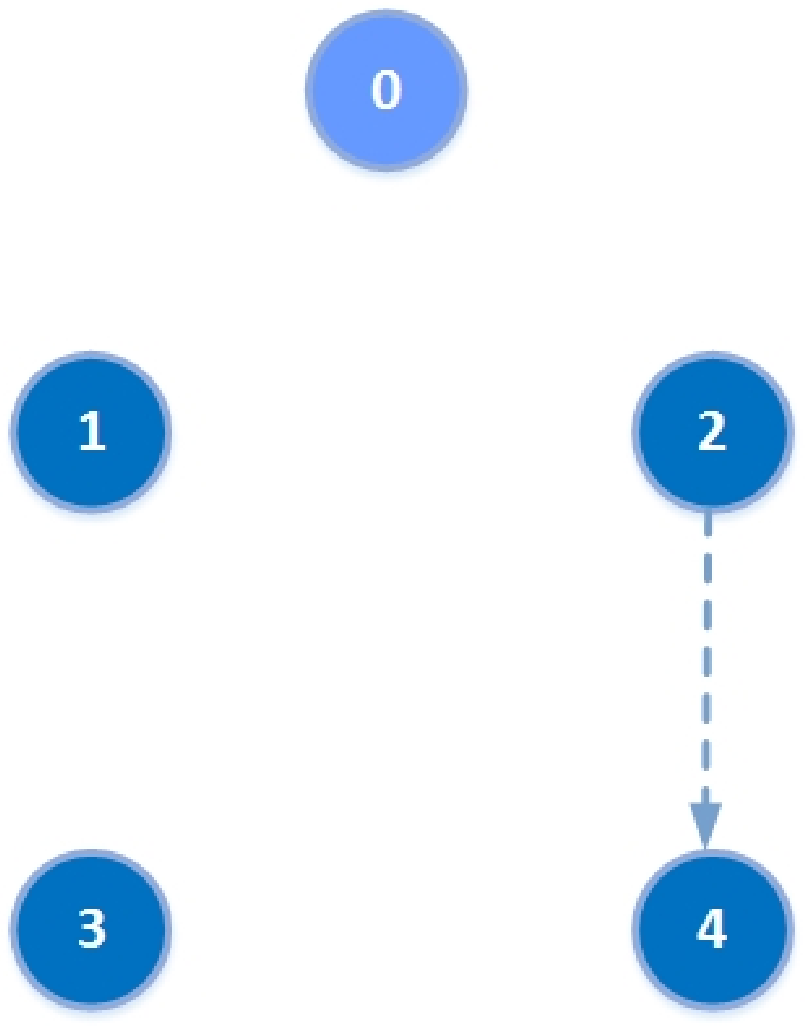}}
  \caption{Switching digraph $\bar{\mathcal{G}}_{\sigma(t)}$ with
$\mathcal{P}=\{1,2,3,4\}$}
  \label{Fig-communication-network}
\end{figure}

The switching digraph $\bar{\mathcal{G}}_{\sigma(t)}$ is described in Figure \ref{Fig-communication-network}
and is dictated by the following switching signal:
\begin{equation*}
   \sigma(t)=
\begin{cases}
1, & \textrm{if}\quad t=8s+0  \ \textrm{or} \  8s+1 \\
2, & \textrm{if}\quad t=8s+2  \ \textrm{or} \  8s+3 \\
3, & \textrm{if}\quad t=8s+4  \ \textrm{or} \  8s+5  \\
4, & \textrm{if}\quad t=8s+6  \ \textrm{or} \  8s+7
\end{cases}
\end{equation*}
where $s =0,1,2,\ldots$.

\begin{figure}
  \centering
  \includegraphics[scale=0.3]{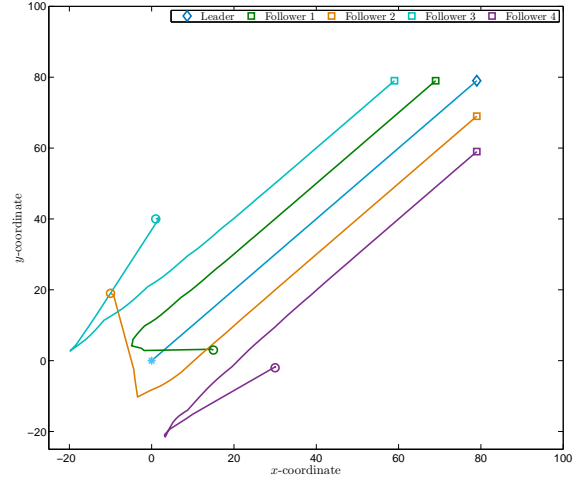}\\
  \caption{The leader-following formation of mobile robots}\label{Figure-trajectory}
\end{figure}

It can be easily verified that Assumptions \ref{ass-jointly-connected} to \ref{ass-solution-to-regulator-equations}
are all satisfied. Therefore, by Theorems \ref{Theo-certain-state-feedback},
this leader-following formation problem can be solved by
a distributed control law of the form (\ref{eq-overall-control-law}).
Simulation of the closed-loop system is performed with
$K_{xi}=\left[
                 \begin{array}{cc}
                   -0.7 & -1.9 \\
                 \end{array}
               \right]\otimes I_{2}
, i = 1,2,3,4$,
 $(x_{d1}, y_{d1})=(-10,0)$,
 $(x_{d2}, y_{d2})=(0,-10)$,
 $(x_{d3}, y_{d3})=(-20,0)$,
 $(x_{d4}, y_{d4})=(0,-20)$,
$v (0) = [0,0,1,1]^{T}$,
$(x_{1}(0),y_{1}(0))=(15, 3)$,
$(x_{2}(0),y_{2}(0))=(-10,19)$,
$(x_{3}(0),y_{3}(0))=(1,40)$,
$(x_{4}(0),y_{4}(0))=(30,-2)$,
and other randomly generated initial conditions.
We let $a_{ij}(t)=1,i,j=0,1,2,3,4,$
whenever $(j,i) \in \bar{\mathcal{E}}_{\sigma(t)}$.
The trajectories of the mobile robots are shown in Figure \ref{Figure-trajectory}.

\begin{Remark}
In this example, the leader's system matrix is not marginally stable.
Thus, as explained in Remark \ref{Remark-advantages}, the approach of \cite{YanHuang17}
is not applicable.
\end{Remark}

%%%%%%%%%%%%%%%%%%%%%%%%%%%%%%%%%%%%%%%%%%%%%%%%%%%%%%%%%%%%%%%%%%%%%%%%%%%%%%%%%%%%%%%%%%%%%%%%%%%%%%%%%%%%%%%%%%%%%%%%%%%%%%%%%%%%%%%%%%%%%%%%%%%%%%%%%%%%%%%%%%%%%%%%%%%%%%%%%%%%%%%%%%%%%%%%%%

\section{Conclusion}\label{Section Six}
In this paper, we have developed a discrete-time distributed observer
and a discrete-time adaptive distributed observer over jointly connected switching networks.
By employing the discrete-time distributed observer,
we have solved the cooperative output regulation problem of a discrete-time linear multi-agent system
over jointly connected switching networks.

\section*{Appendix}
A digraph $\mathcal{G}=\left(\mathcal{V}, \mathcal{E}\right)$ consists of a finite set of nodes $\mathcal{V}=\{1, \ldots, N\}$
and an edge set $\mathcal{E} \subseteq \mathcal{V} \times \mathcal{V}$.
An edge of $\mathcal{E}$ from the node $j$ to the node $i$
is denoted by $(j,i)$, and the node $j$ is called a neighbor of the node $i$.
Then, $\mathcal{N}_{i}=\left\{j \in \mathcal{V}\ | \ (j,i) \in \mathcal{E} \right\}$ is called the neighbor set of the node $i$.
The edge $(i,j)$ is called undirected if $(i,j) \in \mathcal{E}$ implies $(j,i) \in \mathcal{E}$.
The digraph $\mathcal{G}$ is called undirected if every edge in $\mathcal{E}$ is undirected.
If the digraph contains a set of edges of the form $\{(i_{1},i_{2})$, $(i_{2},i_{3})$,
$\ldots$, $(i_{k-1},i_{k})\}$, then this set is called a directed path of $\mathcal{G}$
from the node $i_{1}$ to the node $i_{k}$, and the node $i_{k}$ is said to be reachable from the node $i_{1}$.
A digraph $\mathcal{G}_{s}=(\mathcal{V}_{s}, \mathcal{E}_{s})$
is a subgraph of $\mathcal{G}=\left(\mathcal{V}, \mathcal{E}\right)$
if $\mathcal{V}_{s} \subseteq \mathcal{V}$ and
$\mathcal{E}_{s} \subset \mathcal{E}\cap(\mathcal{V}_{s} \times \mathcal{V}_{s})$.

The weighted adjacency matrix of a digraph $\mathcal{G}$ is a nonnegative matrix
$\mathcal{A}=[a_{ij}]_{i,j=1}^{N} \in \RR^{N \times N}$, where
$a_{ii}=0$ and $a_{ij}>0, i\ne j$, if and only if $ (j,i)\in \mathcal{E}$.
Given a set of $n_{0}$ digraphs $\left\{ \mathcal{G}_{i}=(\mathcal{V}, \mathcal{E}_{i}), i=1,\ldots,n_{0} \right\}$, the digraph
$\mathcal{G}=\left(\mathcal{V}, \mathcal{E}\right)$ with
$\mathcal{E}= \bigcup_{i=1}^{n_{0}} \mathcal{E}_{i}$ is called the union of the digraphs $\mathcal{G}_{i}$,
denoted by $\mathcal{G}=\bigcup_{i=1}^{n_{0}} \mathcal{G}_{i}$.

We call a time function $\sigma: \ZZ^{+} \mapsto \mathcal{P}=\{1,2,\ldots,n_{0}\}$ a piecewise constant switching signal if there exists a sequence
$\{t_{j}, j=0,1,2,\ldots \}$ satisfying $t_{0}=0, t_{j+1}-t_{j}\ge d$ for some positive integer $d$
such that, for all $t\in [t_{j},t_{j+1})$, $\sigma(t)=p$ for some $p \in \mathcal{P}$.
$n_{0}$ is some positive integer,
$\mathcal{P}$ is called the switching index set,
$t_{j}$ is called the switching instant, and $d$ is called the dwell time.
Given $\sigma(t)$ and a set of  digraphs
$\mathcal{G}_{i}=(\mathcal{V}, \mathcal{E}_{i})$, $i=1,\ldots,n_{0}$, with the corresponding weighted adjacency matrices denoted by $\mathcal{A}_{i}$,
$i=1, \ldots, n_{0}$, we call the time-varying digraph $\mathcal{G}_{\sigma(t)}=\left(\mathcal{V},\mathcal{E}_{\sigma(t)}\right)$ a switching digraph, and denote
its weighted adjacency matrix by $\mathcal{A}_{\sigma(t)}$.

\end{document}